# Addressing the Accuracy-Cost Tradeoff in Material Property Prediction: A Teacher-Student Strategy

Dong Zhu[1,2*], Xiaoyu Yang [1,2]
1 Computer Network Information Centre, Chinese Academy of Sciences, Beijing, 100190
2 University of Chinese Academy of Sciences, Beijing

## Abstract

Deep learning has revolutionized the process of new material discovery, with state-of-the-art models now able to predict material properties based solely on chemical compositions, thus eliminating the necessity for material structures. However, this cost-effective method has led to a trade-off in model accuracy. Specifically, the accuracy of Chemical Composition-based Property Prediction Models (CPMs) significantly lags behind that of Structure-based Property Prediction Models (SPMs). To tackle this challenge, we propose an innovative Teacher-Student (T-S) strategy, where a pre-trained SPM serves as the 'teacher' to enhance the accuracy of the CPM. Leveraging the T-S strategy, T-S CrabNet has risen to become the most accurate model among current CPMs. Initially, we demonstrated the universality of this strategy. On the Materials Project (MP) and Jarvis datasets, we validated the effectiveness of the T-S strategy in boosting the accuracy of CPMs with two distinct network structures, namely CrabNet and Roost. This led to CrabNet, under the guidance of the T-S strategy, emerging as the most accurate model among the current CPMs. Moreover, this strategy shows remarkable efficacy in small datasets. When predicting the formation energy on a small MP dataset comprising merely 5% of the samples, the T-S strategy boosted CrabNet's accuracy by 37.1%, exceeding the enhancement effect of the T-S strategy on the whole dataset.

## Introduction

Machine learning methods have risen to prominence as a cost-effective and efficient avenue for new material exploration, catalyzing advancements across diverse fields[1-6], including high entropy alloys[1], clean energy[7], and battery materials[8]. When it comes to exploration costs, models that predict material properties based on structure prove less costly than traditional trial-and-error experiments, and those that predict properties based on composition are even more cost-effective than their structure-based counterparts[9-12]. However, with the decrease in costs comes a gradual diminution in exploration precision, potentially leading researchers down the path of exploring valueless materials[13,14]. The question then arises: ***how can we enhance the precision of exploration while minimizing costs as much as possible?***

There has been a wealth of research utilizing material structure or composition to predict material properties[2-6], which has significantly cut down the cost of exploring the chemical space of materials. In the realm of SPMs, numerous studies have been conducted to fully exploit structure-property knowledge using different model structures or feature representations. This encompasses efforts like CGCNN[15], SchNet[16], ALIGNN[13], and coGN[17], which have progressively minimized the error in predicting material properties. Similarly, in the field of CPMs, there is an abundance of commendable work steadily reducing prediction errors. Works such as ElemNet[18], Roost[19], and CrabNet[14], which employ dense weighting or attention mechanisms between elements, have also achieved remarkable prediction accuracy. Within these two types of property prediction models, the material structure inherently includes the composition details, and the rest of the structure knowledge further bolsters the precision of SPMs. Consequently, the accuracy of CPMs consistently lags behind that of SPMs.

However, the task that presents a greater challenge and holds more value is the exploration of CPMs that are both cost-effective and highly accurate. CPMs merely require chemical formulas as input, whereas obtaining the material structure necessary for SPM prediction involves substantial costs through experiments or theoretical calculations. We might need to strike a balance between accuracy and cost, but the prospect of training CPMs that are both cost-effective and highly accurate is undeniably more appealing.

Drawing inspiration from knowledge transfer[20] and the relationship between material composition-structure-properties, we devised a Teacher-Student strategy. By leveraging the embeddings output of the teacher model (SPM), knowledge is transferred to the student model (CPM), thereby establishing a connection between the composition, structure, and properties of materials. Unlike strategies that employ new model architectures or feature engineering methods, our approach can always enhance the accuracy of CPMs based on existing foundations. As depicted in Figure 1a, we initially pre-train to align the end embeddings of CPM and SPM, compelling the CPM to learn the corresponding composition-structure-property mapping relationship during the training process. Subsequently, the pre-trained CPM is fine-tuned to predict material properties. The advantages of the T-S strategy are quite evident:

The T-S strategy **operates independently of the model structure.** The T-S strategy merely requires the alignment of the end embeddings, hence it is not confined to a specific model structure and can universally enhance CPM's accuracy.

T-S is particularly **well-suited for small datasets in the material field**. The less data available, the less knowledge there is. For CPMs under small samples, the knowledge gleaned from SPMs is more valuable, thus it can significantly improve prediction accuracy.

Leveraging the Teacher-Student (T-S) strategy, we have attained the highest precision in CPMs. On the MP[21] formation energy test dataset, the T-S CrabNet has a prediction Mean Absolute Error (MAE) of 0.048, surpassing the previous benchmark of 0.058, with a relative accuracy improvement of 16.9%. Simultaneously, on the 5% MP dataset, the T-S strategy reduced CrabNet's MAE for formation energy from 0.139 to 0.088, with a relative accuracy improvement of 37.1%. Moreover, we've witnessed considerable advancements in other tasks, achieving the best accuracy performance to date.

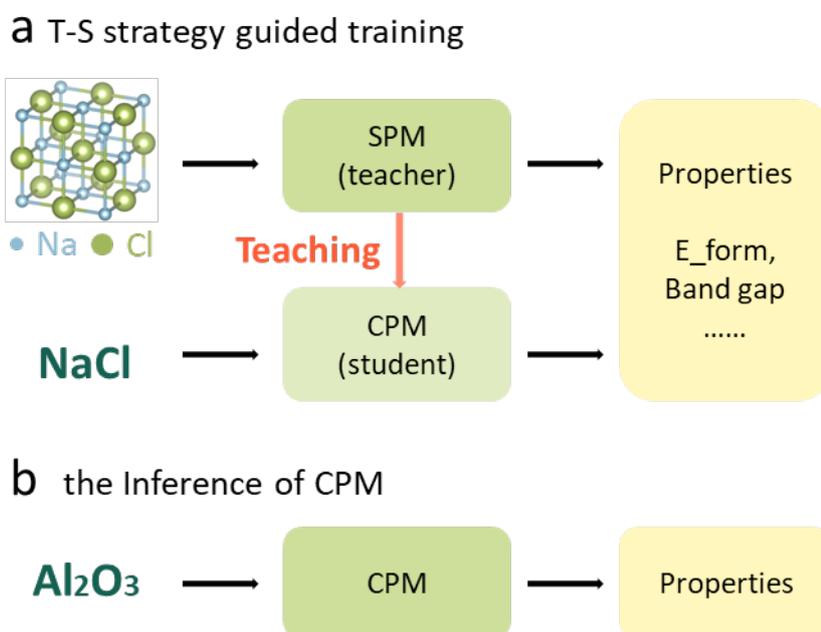

Figure 1. **The comprehensive framework for achieving a low-cost, high-accuracy CPM. a** illustrates the training process of the T-S strategy-based CPM. The SPM, trained with the structure of the material as input, serves as a teacher to train the student CPM, which only takes a chemical composition as input. **b** depicts the inference process of the CPM. Notably, neither the SPM nor the structures of the materials are required for property prediction in the inference process.

## Related Work

Pre-training and Fine-tuning Paradigm

The pre-training and fine-tuning paradigm has been making waves in the Natural Language Processing (NLP) and Computer Vision (CV) fields in recent years. This two-step dance starts with models cutting their teeth on a vast playground of unlabeled data, picking up universal representations. Then, they hone these skills on task-specific labeled data, adapting to the nuances of specific applications. BERT[22], in the NLP sphere, leverages the Transformer architecture to pre-train on a sea of text data, delving into the depths of language representations. Hot on its heels came larger models like GPT-3[23], a behemoth with 175 billion parameters, flexing its muscles across a spectrum of language tasks. Successors like ChatGPT and GPT-4[24] have continued to underscore the allure of the pre-training and fine-tuning paradigm, coupled with large-scale models. In the field of Computer Vision, the pre-training and fine-tuning paradigm has also found widespread application. Models like MAE[25], which harness this paradigm, and the SAM[26] model have been turning heads with their impressive results. Our approach, however, takes a different tack from this unsupervised data-based pre-training. We nudge our CPM to learn from the SPM's embeddings during pre-training, thereby teasing out latent representations of the composition-structure-property triad. The fine-tuning phase is not limited to supervised learning; instead, it continues under the guidance of SPM to fine-tune the model for optimal performance.

Knowledge Distillation

Knowledge distillation is an emerging strategy aimed at constructing efficient small-scale networks[27]. The central idea is to transfer the "knowledge" from complex teacher models into simpler student models. This concept, initially brought to light by Hinton[28] and his colleagues, primarily involves training a smaller model to emulate the behavior of its larger counterpart. The main advantage of this strategy is its ability to reduce model complexity and computational requirements while maintaining high performance. In the wake of this, various knowledge distillation methods have been applied across different fields with the goal of maximizing model performance while minimizing model size[29-31]. Furthermore, cross-modal knowledge distillation models have emerged, where the inputs for the teacher and student models are entirely different, yet share an underlying semantic connection[32-34].

Our strategy, while drawing inspiration from these works, is distinct in its approach. The key to our strategy is that the material structure input into the SPM contains the composition information input into the CPM, with the composition also having the capability to infer the structure to a certain degree. This inclusion yet interrelation of input information makes past strategies no longer applicable. The volume of input information results in inconsistent upper limits of model accuracy, but the interrelation between composition and structure also offers a theoretical avenue for enhancing the performance of weaker models.

## T-S strategy

**Two step strategy**

At the heart of our T-S strategy is the transfer of knowledge from SPMs to CPMs. Inspired by the pre-training-fine-tuning model and distillation learning, we divide the T-S strategy into two distinct steps: Knowledge Transfer and Knowledge Guidance.

Step 1. **Knowledge Transfer** involves utilizing a pre-trained SPM as a 'teacher' to impart knowledge to a 'fresh' CPM. As shown in step 1 of Figure 2, the CPM's task during this phase is to align its generated embeddings with those produced by the SPM, thereby facilitating the CPM's understanding of the relationships between composition, structure, and properties. The input of the SPM is the structure corresponding to the components of the CPM model input, ensuring the correctness of the relationship. Throughout this process, the CPM does not predict the target properties, so the final linear layer is not involved in the update of parameters. The loss function for Step 1 is be defined as:

$$L_{step1} = L_{embedding} = MSEloss(Embedding_{SPM}, Embedding_{CPM})$$

Here, $Embedding_{SPM}$ and $Embedding_{CPM}$ refer to the embeddings output by the network just before the final linear layer of the SPM and CPM, respectively. This approach ensures the effective transfer of knowledge from the SPM while preserving the flexibility of the SPM and CPM model structures.

Step 2. **Knowledge Guidance.** During the Knowledge Guidance phase, the SPM, acting as a teacher, transitions from being a knowledge instiller to a guide, steering the CPM towards a more accurate path. As shown in step 2 of Figure 2, the SPM acts as a knowledgeable guide in this stage. The ultimate goal for the CPM, as a student, is to predict the properties as accurately as possible. Therefore, we introduce a new loss, the actual prediction loss, to guide the CPM's parameter update.

$$L_{step2} = \frac{alpha * loss\_embedding + beta * loss\_target}{alpha + beta}$$

We set beta to 1 by default, and by adjusting the value of beta, we can control the proportion of the embedding loss in the overall loss.

In the two-step process, we can use either the same dataset (D1) or different datasets (D1 and D2). D2 is the dataset we want to train on, while D1 can include larger datasets like MP and JARVIS[35]. By using these larger datasets, we can transfer more generalized knowledge about the composition-structure-property relationship.

In the actual training process, Step 1 can be seen as a special case of Step 2. When beta is set to 0, Step 2 becomes Step 1, making the transition between the two training stages straightforward and without additional workload. Both stages contribute to some performance improvement for the CPM. If only Step 2 is used, in the prediction of formation energy on the whole mp database, CrabNet (CPM) reduces the MAE from 0.058 to 0.051. With the addition of Step 1, the MAE is further reduced to 0.048.

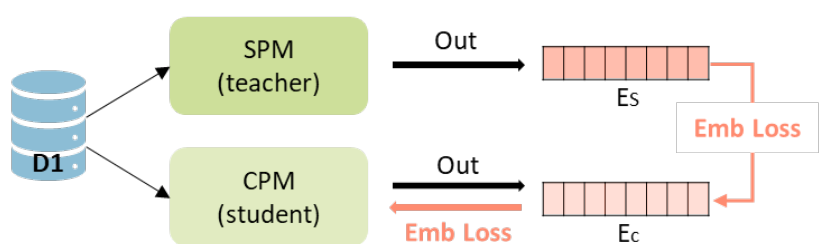
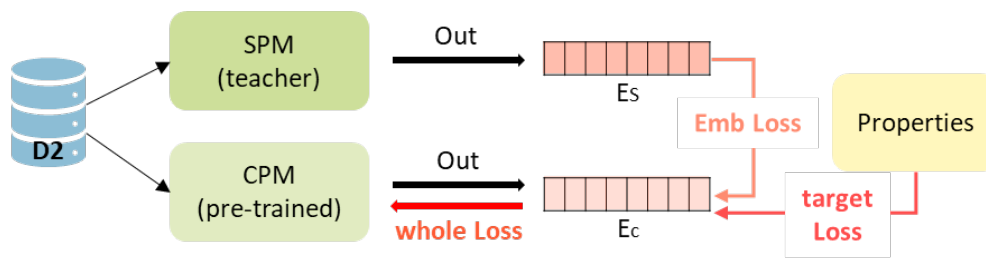

Figure 2. **Implementation process of the T-S strategy.** Step 1 represents the knowledge transfer

phase, while Step 2 signifies the knowledge guidance phase. Both stages contribute significantly to the enhancement of the CPM's performance in predicting properties using the chemical compositions.

**Implementation of the T-S Strategy**

In the T-S strategy, we employ a pre-trained ALIGNN as the seasoned teacher, while the untrained CrabNet and Roost take on the roles of eager students. The T-S strategy is universal for different SPMs and CPMs, but for the sake of clarity, we'll use ALIGNN and CrabNet as our examples to illustrate the finer points of the implementation process. At the same time, we control the training parameters to be the same as the source papers.

The architectural design of ALIGNN stays true to its original form. The only adjustment in the code is that the output of the final layer is now 256-dimensional embeddings, tailored for the T-S strategy. As for CrabNet, the main body of the code remains unchanged. H However, to align with the embedding dimension output by ALIGNN, we have tweaked the size of the final ResNet layer in CrabNet from [256, 128, 64] to [256, 256, 256] (this adjustment isn't mandatory, as demonstrated by the unmodified Roost). In step1, we're solely focused on calculating the loss of the embeddings output by these two models. Then, in the second step, we broaden our scope to calculate both the embedding loss and the target loss.

# Experiments

**University of the T-S strategy**

On the Materials Project and Jarvis database, our T-S strategy has demonstrated a range of enhancements in the performance of CPMs for material property predictions. In the course of this experiment, we maintain a consistent dataset for both Step 1 and Step 2 within the T-S strategy. This substantiates the notion that the Teacher-Student strategy can universally elevate the performance ceiling of CPMs, unrestricted by any specific task. To authenticate the model-agnostic nature of the T-S strategy, we examined the enhancement capability of ALIGNN, serving as the teacher model, on two student models, namely CrabNet and Roost. As depicted in Table XX, our experiments revealed that CPMs benefit from the guidance of the T-S strategy on the formation energy and band gap prediction tasks across the MP and Jarvis datasets. Nonetheless, Roost's Jarvis band gap prediction task did not witness any improvement in model accuracy, while CrabNet exhibited an enhancement. We infer that this is attributable to Roost's architecture's limited capacity for learning composition-structure associations, resulting in a virtually negligible improvement in this task.

| Dataset | MP_e_form | MP_band_gap | Jarvis_e_form | Jarvis_band_gap |
|---|---|---|---|---|
| CrabNet | 0.058 | 0.366 | 0.088 | 0.200 |
| **T-S CrabNet** | **0.048** | **0.327** | **0.084** | **0.183** |
| Roost | 0.060 | 0.387 | 0.103 | 0.210 |
| **T-S Roost** | **0.054** | **0.362** | **0.095** | 0.210 |
| ALIGNN (teacher) | 0.022 | 0.218 | 0.033 | 0.140 |

Table 1. **Performance of CPMs (CrabNet and Roost) and the ALIGNN model on the formation energy and band gap prediction tasks in the MP and Jarvis datasets**. The prefix "T-S" indicates that the model has been aided by the Teacher-Student (T-S) strategy. The values represent the Mean Absolute Error (MAE) of the models on the test set. It is evident that the T-S strategy reduced the MAE of the CPMs.

**The performance of the T-S strategy on small datasets**

The T-S strategy demonstrates a significant performance boost when applied to small datasets. The trend we observed is that the smaller the dataset, the more pronounced the improvement achieved by the T-S strategy. Considering the high cost of data acquisition in the materials science field, this trend is especially advantageous for small datasets typical of this domain, helping to mitigate the pressure from the extensive data demands of deep learning.

In order to prevent a decrease in performance caused by the inability to utilize a dataset that is 100% complete in step 1, we employ separate datasets for both step 1 and step 2 in T-S strategy. For instance, we utilize the MP dataset in step 1 and the JARVIS dataset in step 2 when training on the JARVIS dataset.

We carried out tests to assess the enhancement effect of the T-S strategy on CPMs across different dataset sizes: 5%, 25%, 50%, and 75%. As shown in Figure 3, the smaller the dataset, the more substantial the improvement effect of the T-S strategy. we evaluate the performance of the T-S strategy on CrabNet across four tasks, specifically on the MP and JARVIS datasets. The most significant improvement was seen in the 5% dataset, where the T-S strategy reduced CrabNet's MAE on the MP formation energy test set from 0.139 to 0.088. This MAE performance nearly matches the original performance of CrabNet on a dataset five times larger (25%). The smaller the dataset of chemical composition-property pairs, the less knowledge it can offer to the deep learning model. At the same time, the guidance from the knowledge-rich SPM becomes increasingly vital during the training of CPMs. This is likely why the T-S strategy yields more benefits for CPMs training from small datasets.

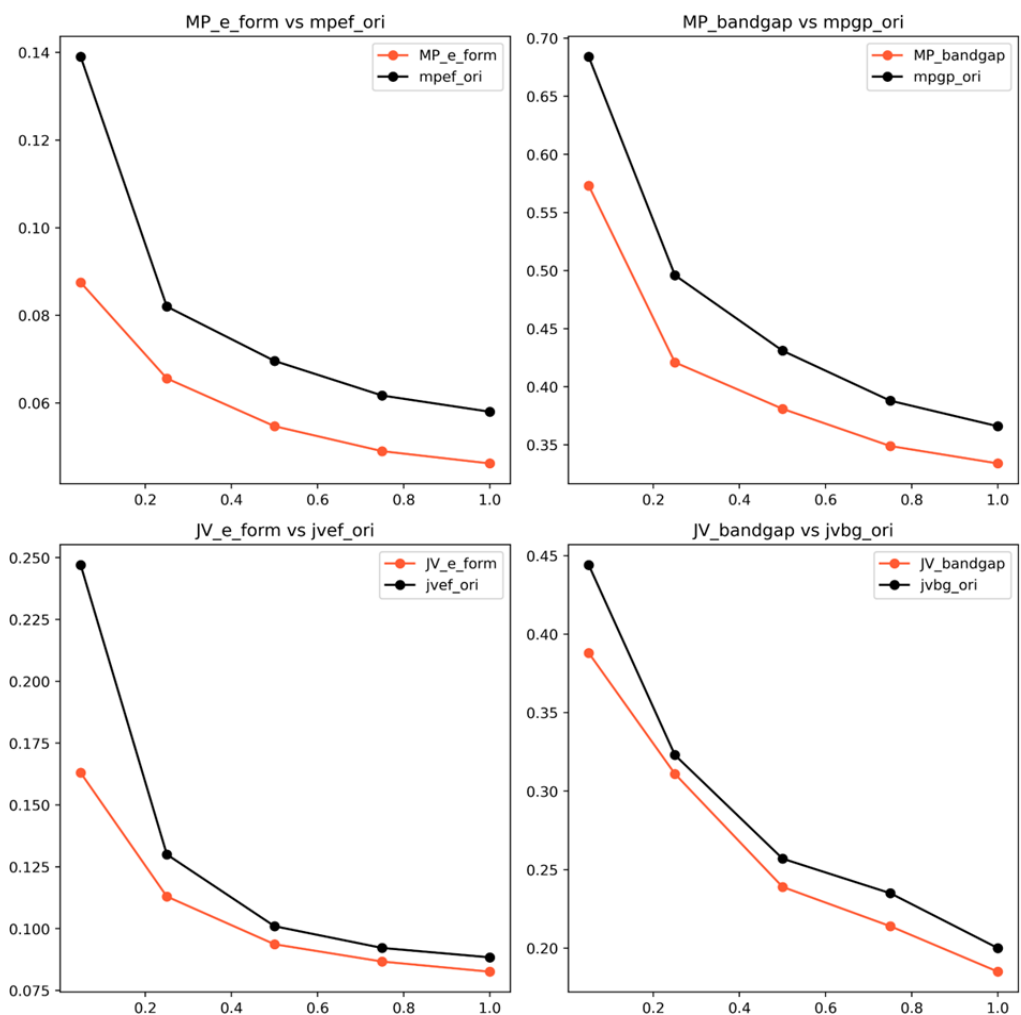

Figure 3. **Performance of the CrabNet model with and without the T-S strategy** on the four tasks for the MP and JARVIS dataset at varying data volumes (5%, 25%, 50%, 75%, and 100%). The red line represents the performance of CrabNet when the T-S strategy is employed, while the **black** line indicates the performance of the original CrabNet model. The top left graph represents the MP formation energy task, the top right graph the MP band gap task, the bottom left graph the JARVIS formation energy task, and the bottom right graph the JARVIS band gap task. The T-S strategy significantly reduces the Mean Absolute Error (MAE) of the CPM. Notably, the relative improvement in precision brought about by the T-S strategy increases as the volume of data decreases.

**Gain Produced by Step1 Step1 in T-S Strategy**

We validated the crucial role of step1 within the T-S strategy by executing tests across four tasks with a data volume of 5%. The removal of step1 led to a substantial reduction in the efficacy of the T-S strategy. The values presented in the table correspond to the Mean Absolute Error (MAE).

|            | T-S strategy | No step1 |
|------------|--------------|----------|
| MP_e_form  | **0.0875**   | 0.144    |
| MP_bandgap | **0.573**    | 0.618    |
| JV_e_form  | **0.163**    | 0.236    |
| JV_bandgap | **0.388**    | 0.445    |

Table xx. Table xx. The performance of the model with and without the use of step1 in the T-S strategy. Step1 can significantly enhance the performance of the model.

**The Impact of Variations in Embedding Loss Proportion on Performance**

Loss Function: $loss = \frac{alpha*loss\_embedding + beta*loss\_target}{alpha+beta}$

The extent of guidance from SPMs significantly influences the ultimate enhancement of the CPM. Taking the prediction of formation energy on the MP dataset using CrabNet as an example, Figure 4 demonstrates that incorporating embedding loss results in a substantial uplift in CPM performance when guided by a teacher model, compared to the special case where alpha equals zero. In our quest to understand the performance enhancement attributed to the level of guidance in the Teacher-Student strategy, we conducted tests to assess the influence of varying embedding loss ratios on model performance. In the loss function, we fixed beta at 1 and varied alpha among 0, 20, and 5 (0 signifies the special condition of absence of guidance).

Naturally, for diverse tasks, the absolute values of embedding loss and target loss will fluctuate, leading to changes in their relative ratios. As a rule of thumb, fine-tuning alpha to approximate the values of the two losses often yields superior model performance.

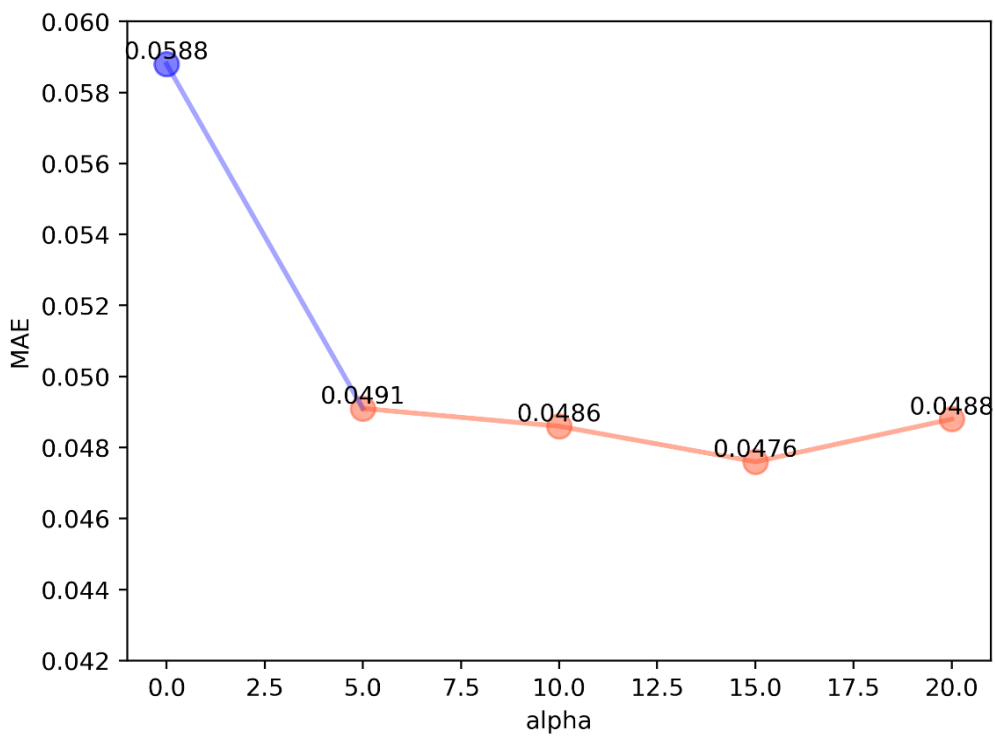

Figure 4. **The influence of varying alpha values in step2 on the model's performance.** An alpha value of 0 signifies the absence of embedding loss. The introduction of embedding loss emerges as a critical factor in enhancing model performance. When alpha is 0, the model's MAE is significantly higher than when embedding loss is incorporated.

**Changes Induced by T-S Strategy in CPM Embeddings**

We visualized the role of the T-S strategy during the training process of CPM. As illustrated in Figure 5, we utilized the T-SNE[36] technique for dimensionality reduction and visualization of the final embeddings layer of Crabnet and ALIGNN in the context of the formation energy task on the MP dataset. It becomes clear that ALIGNN's embeddings exhibit a superior aggregation for materials with proximate property values. Contrasting with the embeddings derived from Crabnet, ALIGNN effectively separates materials with significant property differences, avoiding the blending of materials with large property disparities. This indirectly infers a higher model accuracy for ALIGNN over CrabNet.

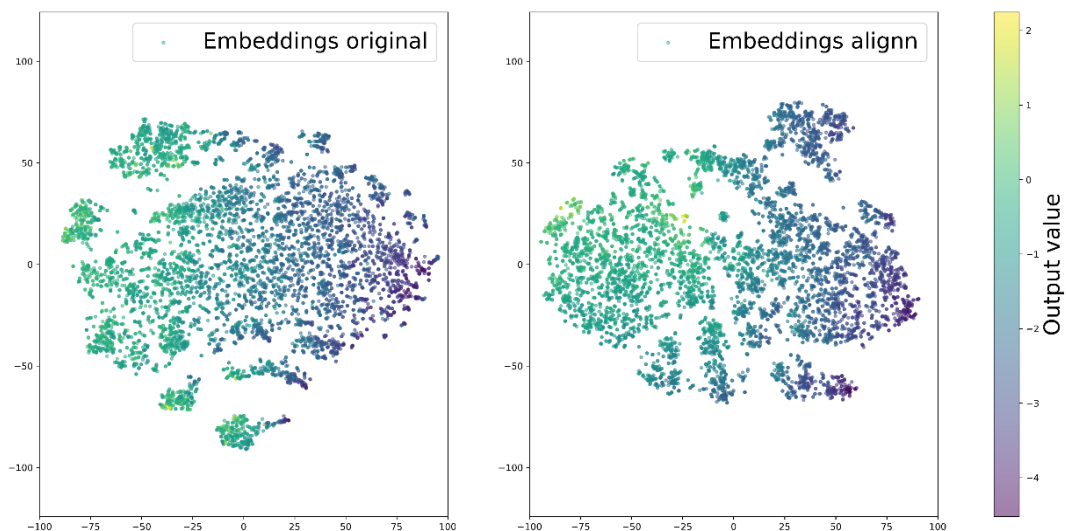

Figure 5. **A two-dimensional visualization comparison of the final embeddings of the original CrabNet (CPM) and ALIGNN (SPM).** The left side of the visualization represents the embedding of CrabNet, while the right side represents the embedding of ALIGNN. The CrabNet embedding tends to mix materials with different properties in the middle area, whereas ALIGNN effectively distinguishes and separates them.

Additionally, we evaluated the disparities between the ALIGNN embeddings and the original CrabNet, as well as the disparities between T-S CrabNet's embeddings and original CrabNet. Observing Figure 6, it is clear that the distribution of the two is highly similar. This potentially indicates that the T-S strategy facilitates CPM in assimilating the additional knowledge encapsulated in SPM. From a materials science perspective, it can be posited that the T-S strategy imparts to CPM the absent composition-structure relationship, thereby establishing the composition-structure-property relationship and augmenting the accuracy of CPM in predicting properties based on compositions.

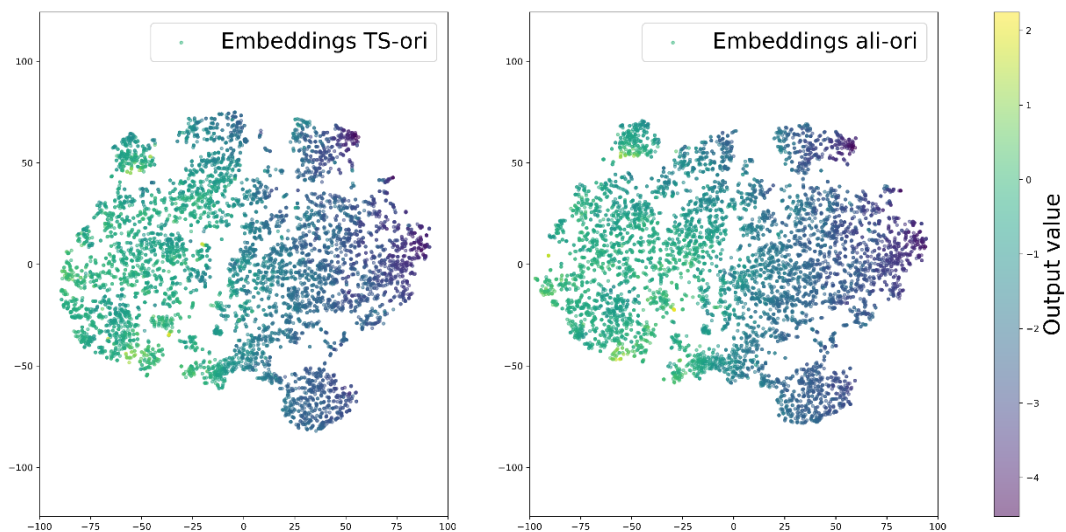

Figure 6. **A visual demonstration of the impact guided by the T-S strategy.** The left side of the image displays the disparity between CrabNet's embedding with and without the T-S strategy, highlighting the effect brought about by the T-S strategy. On the right side, we observe the difference in embeddings between ALIGNN and CrabNet(without T-S strategy). The two-dimensional visualization of both sets of embeddings reveals a striking similarity in the degree of separation, size, and shape of each cluster, as well as the effective separation of parts with significant property differences. This suggests that the T-S strategy enables CrabNet to learn a distribution of embeddings that closely resembles that of ALIGNN, thereby indirectly demonstrating that the T-S strategy assists CPM in establishing the mapping relationship between composition, structure, and property.

## Conclusion

We successfully elevated the precision of cost-effective CPMs with our proposed Teacher-Student (T-S) strategy. By employing ALIGNN from the SPM as the teacher and CrabNet and Roost from the CPM as students, we showcased the advantages of our strategy on the MP and Jarvis datasets. The SPM has facilitated the CPMs (CrabNet and Roost) in forging a component-structure-property relationship, thereby enhancing the model's precision and validating the versatility of the T-S strategy. In addition, we executed experiments on the MP and JARVIS datasets with diverse data volumes. We discovered that the smaller the dataset, the more significant the relative accuracy improvement yielded by the T-S strategy. This indicates that the T-S strategy can offer substantial benefits to deep learning models operating under limited data conditions. Naturally, there are numerous potential avenues for future exploration of this T-S strategy, including experimenting with alternative methods to transfer the knowledge from SPM to CPM.